# Converting water adsorption and capillary condensation in useable forces with simple porous inorganic thin films

Mickael Boudot[1], Hervé Elettro[3], David Grosso[1,2]*

[1] Sorbonne Universités, UPMC Univ Paris 06, CNRS, Collège de France, UMR 7574, Chimie de le Matière Condensée de Paris, F-75005, Paris, France.
[2] NSE-IM2NP - UMR CNRS 7334, Aix-Marseille Université, Faculté des Sciences de Saint Jérôme, Case 142, 13397 Marseille Cedex 20, France
[3] Sorbonne Universités, UPMC Univ Paris 06, CNRS, UMR 7190, Institut Jean Le Rond d'Alembert, F-75005 Paris, France

**Abstract**.
This work reports an innovative humidity driven actuation concept based on Bangham effect using simple nanoporous sol-gel silica thin films as humidity responsive materials. Bilayer shaped actuators, consisting on a humidity-sensitive active nanostructured silica film deposited on a polymeric substrate (Kapton) were demonstrated as an original mean to convert water molecule adsorption and capillary condensation in useable mechanical work. Reversible silica surface energy modifications by water adsorption and the energy produced by the rigid silica film contraction, induced by water capillary condensation in mesopores, were finely controlled and used as the energy sources. The influence of the film nanostructure (microporosity, mesoporosity) and thickness, and of the polymeric support thickness, on the actuation force, on the movement speed, and on the amplitude of displacement are clearly evidenced and discussed. We show that the global mechanical response of such silica-based actuators can be easily adjusted to fabricate a humidity variation triggered tailor-made actuation systems. This first insight in hard ceramic stimulus responsive materials may open the door toward new generation of surface chemistry driven actuation systems.

**Introduction.**
Adaptive material or system able to produce a mechanical work in response to a specific external stimulus or a change in their environment is the general definition that describes an actuator. Nowadays such materials or systems arouse a great interest in many fields as different as materials science, medicine, engineering, robotics, etc.[1,2]
Man-made actuators concepts rely most of the time on synthetic responsive materials able to convert an energy source in a controlled movement. A wide range of organic and inorganic materials were developed to respond to various types of energy-giving stimuli. For instance, electro-active polymers[3,4], carbon nanotubes based structures[5,6], piezoceramics[7,8] or metallic electrochemical thin films[9,10] ,to name a few, which are studied for their response to electric stimuli. Thermo-responsive elastomers[11], shape-memory ceramics[12,13] and alloys[14] are known for their ability to convert thermal energy in mechanical work. Concepts that exploit magnetic and light responsive materials, belonging to the category of pure man-inspired actuators, are also interesting.[15,16] Another approach, which is inspired by biological systems, consist in mechanical movements generated from chemical energy sources as observed with plants for example.[17,18] Synthetic actuation mechanism driven either by gas adsorption, molecule diffusion or pH stimulus were some bio-inspired examples to turn the chemical energy into artificial mechanical movement.[19–22] A fast actuation speed, a large amplitude of movement and a robust reactivity are the essential requirements of the sensitive material to tailor valuable actuators and create various ingenious smart devices as energy generators[19,23], artificial muscles[24] or programmable origami.[25]

Recently nanostructured materials attracted a lot the attention for actuator design. Enhanced performances (energy efficiency, deformation rate, etc.) combined to lower triggering activation energies were observed in particular with surface driven actuation mechanisms that utilize nanostructured (porous) materials, which is mainly due to the nanostructure-induced exalted surface to volume ratios.[26–28] In addition to improving existing actuators technologies, such nano-tailored materials open gates toward new actuation mechanisms. An example is the nanoporous structures that authorized liquid capillary condensation in pores with sizes comprised between 2 nm and 50 nm. Such a phenomenon has recently been suggested, and yet not investigated, as a new actuation concept that would lead to humidity induced actuation devices. Here, "hard" inorganic porous cohesive networks, with the potential generation of high mechanical forces, would be required.[29] For instance, it was recently observed that liquid-gas meniscus forming through capillary adsorption and desorption in mesoporous silica films is accompanied by a significant transversal compressive deformation (≈ 1-10%).[30] The "huge" Laplace forces (up to few GPa) generated within the interconnected network of nanopores associated to mesoporous materials, are responsible for the so-called capillary contraction.[31,32] The harnessing of the Laplace force and its conversion in a macroscopic work seems thus to be a promising way to convert an easy to reach chemical energy in exploitable force. A technological solution would then associate a layer of the stimulus responsive materials onto a passive elastic substrate. This is one of the simplest system that can convert a stimulus-induced volume change of an active material into a macroscopic mechanical work.[33,34] The challenge here consists in elaborating the tailored nanostructured porous active layer onto the proper passive elastically deformable support, insuring a strong and homogeneous bonding between both materials.

In the present article, we demonstrate that such humidity sensible actuators can be easily elaborated by applying sol-gel silica films with controlled nanoporosity on one face only of a thermal resistant polymer sheet, in analogy to the bimetal strip designed to measure temperature.[35] The thermal condensation curing, following the deposition of the silica layer, induces a contraction of the inorganic network and induces a lateral stress into the substrate that provokes the curvature of the whole sheet. The adsorption at the surface of the porous silica layer and the capillary condensation into the nanopores, which are in equilibrium with the applied humidity, create strains inside the whole porous materials that add to the previous stress. The result is the modification of the curvature radius of the sheet with respect to the applied humidity. If the additional strain induces a contraction of the porous network the curvature radius decreases, while if it induces an expansion the radius increases, converting thus the adsorption/desorption and capillary condensation/evaporation in reversible actuation movements. We show that the evolution of the curvature can be more or less complex depending on the type of porosity. The movements and forces generated by the bilayer sheet actuators were assessed combining measurements at nano and macro scale by ellispometry technics, optical observations and micro-force measurements. The relatively high generated forces make this actuation concept an interesting potential water-fuelled solid engine which was illustrated by an autonomously-moving system in following parts.

**Results and Discussion.**

*Fabrication and characterization of the bilayers actuators.*

Bilayer structures are an assembly composed of a silica thin film and a flexible 50 μm thick polyimide as support. Silica coatings were synthesized by classical sol-gel dip-coating methods on polyimide (poly 4,4'-oxydiphenylenepyromellitimide) sheets from hydro-alcoholic solutions composed of TEOS (TetraEthylOrthoSilicate) precursor. Mesoporous thin films were prepared using the Evaporation Induced Self-Assembly (EISA)[36] method in presence of block-copolymer micellar template Pluronic F127, while microporous silica xerogel films were obtained from similar solutions but free of block-copolymers. All films were stabilized by a thermal curing at 350°C for 10 min under air (see Materials and Methods part). Polyimide substrate was chosen because of its chemically inertia, its thermal stability (-269 °C and 400 °C), and its mechanical properties.[37] Bilayer structures are labelled $S^X$ or $S^M$ either they were fabricated using a Xerogel or Mesoporous silica film. It's known that growth of a thin film on unique face of a substrate is susceptible to induce the substrate deformation depending on the condition of the coating formation. Considering a coating with a homogeneous thickness on a plane substrate, this deformation would be spherical and either concave when the mechanical stress created by the coating and applied on the substrate is a compressive stress (σ < 0) or convex when it is a tensile stress (σ > 0). For our silica-polyimide bilayers shown in Figure 1.a, two distinct steps are observed. First, immediately after the deposition of the silica sol gel solution by dip coating, solvents evaporate leading to formation of a thin film composed of silica intermediate moieties. The drying of the sub-micro-meter-thick gel film at room temperature induces total stress in the plane of the coating which is approximately equivalent to the capillary pressure in the liquid.[38] Nevertheless the residual stress generated by the transformation from the liquid film to gel is relatively weak because the loss of thickness of the coating is accommodated by viscous and plastic strains due to the low condensation degree of the intermediate silica matrix. Therefore the bilayer structures do not shown any curvature after the film drying as shown on Figure 1.b.

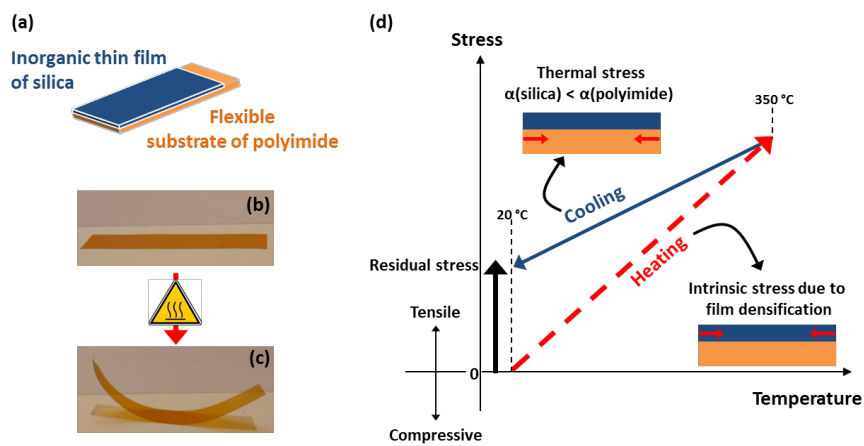

Figure 1. (a) Representation of the silica-polyimide bilayers sheet. (b) Photography of a bilayer sheet after drying at room temperature. (c) Photography of the corresponding bilayer sheet after calcination at 350°C for 10 min. (d) Schematic representation of stress evolution in silica thin films during a thermal run composed of heating and cooling steps, adapted from Ref. 39.

In the second step, the average thickness contractions of 19% and 38% measured by ellipsometry of the xerogel and mesoporous silica thin films respectively is due to the densification of the coatings by condensation of the inorganic network during the thermal curing at 350 °C for 10 min. Due to the thermal-induced silica volume reduction and because the film is relatively strongly bonded to the polyimide substrate, the thermal contraction of films is mainly accommodated by a global film volume decrease that is clearly observed through the thickness reduction. In-plane stresses, parallel to the substrate-coating interface, are then generated and make the bilayers sheet curve as shown in Figure 1.c.[39] The curvature formation in such bilayers systems was explained by Kozuka et al.[40] and is represented in Figure 1.d. To summarize, under heating the condensed coatings can be considered as retained in elongated state by the substrate, the generated stress cannot, or very partially, be dissipated by any viscous or plastics strains. This in-plane tensile stress is called intrinsic stress, Figure 1.d. The intensity of the intrinsic stress and the resulting curvature of the sol-gel/polyimide bilayers depends on the thermal curing conditions, on the cristallinity or the amorphous nature of the coating, and on the hydrolysis ratio of the inorganic solution.[41–43] No significant chemical reaction is expected to occur during cooling to room temperature (20 °C). Nevertheless the drop in temperature is accompanied by a decrease of the bilayer curvatures. Indeed a thermal stress is generated due to the difference of the thermal expansion coefficient (α) between the silica and the polyimide substrate. The lower α value of silica compared to polyimide, $\alpha_{Si}$ = 5.5.10$^{-7}$ °C$^{-1}$ and $\alpha_K$ = 30-60.10$^{-6}$ °C$^{-1}$ respectively induced a compressive stress opposed to the tensile intrinsic stress. This thermal expansion coefficient related stress is negligible during the heating step.[44] As a result, the curvature observed at room temperature corresponds to the residual stress, which is the sum of the intrinsic and thermal stresses. For the sake of simplicity, a rectangular flat geometry has been chosen for the fabrication of the bilayers (L*l = 60*6 mm$^2$). This geometry disfavored the transversal (l-direction) in-plane stress and reduces thus the formation of a curvature along the width of the sample. From now, only longitudinal (L-direction) in-plane stress will be considered. At least 50 nm of coating was required to bend the substrate. The residual in-plane stress in bilayer structures can be estimated by measuring the radius of curvature of the elastic isotropic substrate where the coating is deposited using the following Stoney's equation (Eq. 1).

$$\sigma_f = \frac{E_s h^2}{6 t_f (1-\nu_s) R} \qquad \text{Eq. 1}$$

Where $\sigma_f$, $E_s$, $h$, $t_f$, $\nu_s$ and $R$ are the residual stress in the film, the Young's modulus of the substrate ($E_s$ = 2,5 GPa at 25°C), the thickness of the substrate, the thickness of the silica coating, the Poisson's coefficient of the substrate ($\nu_s$ = 0,3) and the radius of curvature of the bilayer sheet respectively.[45,46]

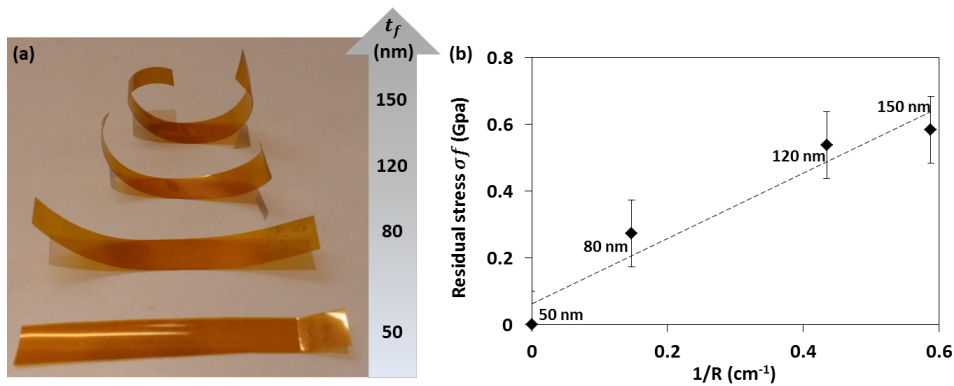

Figure 2. (a) A photography of the curvature of 50 μm-thick polyimide strips combined with a layer of xerogel silica with decreasing thickness from top to bottom, and (b) the graphic of the corresponding calculated residual stress deduced from equation 1.

The residual stress determined by the Stoney's equation increases linearly with the xerogel coating thickness, as reported in Figure 2.b. A value of around 500 MPa was determined for the thicker silica coating (150 nm). In-plane stress values of the same order were determined by Parril et al.[42] $S^X$ and $S^M$ bilayer strips, showing radius of curvatures ($R$) of 5.0 ± 0.3 cm and 6.8 ± 0.3 cm at 10 % of relative humidity, respectively, were obtained with 80 ± 5 nm and 252 ± 5 nm thick xerogel and mesoporous silica thin films, respectively. The higher curvature of $S^X$ compared to $S^M$ despite a significant lower thickness reveals that, in addition to the thin film thickness, the structure of pore network also influences directly the residual stress. This highlights the fact that the in-plane residual stress is related to the density of the porous material but also on the pore characteristics. The smaller the size of the pores and the higher the density of the siliceous network are, the higher is the stress generated during thermal curing. Porosity and pore size distribution of both silica films were assessed by environmental ellipsometry porosimetry. Corresponding water adsorption-desorption isotherms are reported in Figure 3.a and 3.b, for both silica films deposited on plane silicon substrates (100), assuming the structures are the same than on polyimide substrates. The isotherm of the xerogel film shows a water-uptake mainly for relative humidity below 20%, which is characteristic of microporosity. Such xerogel typically contain less than 2 nm in diameter micropores, since the curing temperature of 350°C was not high enough to completely densify the inorganic network. The water adsorption/desorption within the mesoporous silica film shows the expected typical type IV isotherm of interconnected pore network. Pore volumes of 10% and 49% were determined for the xerogel and mesoporous films respectively, using the Bruggeman effective medium approximation.[47] Capillary condensations are observed at water partial vapor pressures of 0.70 for the mesoporous film. The pore size distribution was determined using the Kelvin equation (Eq. 2) for mesoscopic pores:

$$\ln \frac{P}{P_0} = - \frac{2\gamma V_l \cos\theta}{r_p RT} \qquad \text{Eq. 2}$$

where ($P/P_0$) is the relative pressure and $r_p$, $\gamma$, $V_l$, $\theta$, $R$, $T$ are the Kelvin pore radius, the surface tension, the molar volume of liquid, the contact angle, the gas constant and the temperature, respectively. Pore size distributions taken from adsorption curves (Figure 3.b - inset), show Gaussian-like distributions in the range of 4.3-8.1 nm with a corresponding mean pore diameters of 5.5 ± 0.40 nm. Because liquid water / vapour interfaces are suspected to not exist below 2nm, the Kelvin equation is not adapted to micropores of the xerogel film. We thus consider that xerogel films contain micropores below 2 nm in diameter, with a large size distribution. GI-SAXS pattern of the mesoporous silica film is shown in Figure 3.b (inset) corresponds to a typical pseudo-cubic (Im3m) structures.[48]

*From bilayer strips to actuator.*
Actuation works of the silica-polyimide bilayer strips were investigated by fixing the $S^X$ or $S^M$ samples into an environmental chamber with controlled atmosphere at room temperature (see Figure 4.a). The humidity in the chamber was controlled between 10 % and 85 % using a RegulHum flow system from SolGelWay Company. A low flow of 1 L.min$^{-1}$ was used to avoid an oscillating movement of the strips. The photomontages displayed in Figure 4.b show the variation of the strip curvature with respect to the applied humidity for both samples, confirming the efficient conversion of humidity stimuli to macroscopic mechanical movement for the two types of porous structures.

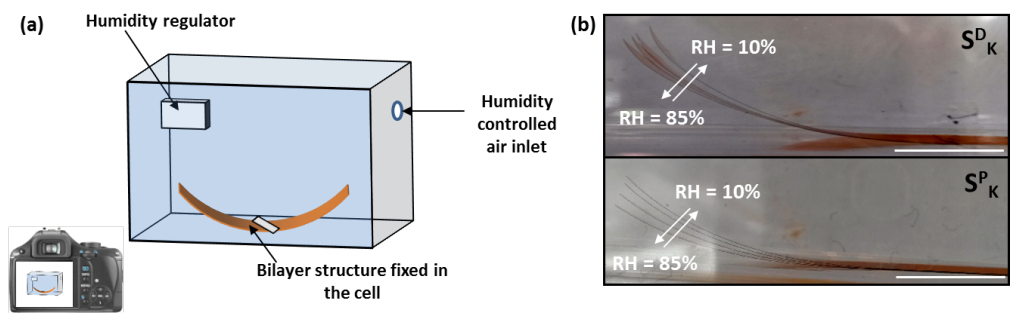

Figure 4. (a) Representation of the set up used for the dynamic *in situ* investigation of the curvature radius depending on the humidity for the S$^X$ and S$^M$ bilayers strips. (b) Photomontages of the S$^X$ and of the S$^M$ bilayers, depending on the relative humidity between 10 % and 85 % at room temperature. Bar scale corresponds to 1 cm.

*Response of the Xerogel microporous thin film:*
To understand the actuation mechanism, the effect of water adsorbed onto the top surface and in the micropores of the xerogel silica has been observed *in situ* by recording the variation of the curvature of the S$^X$ bilayers with the relative humidity in the chamber. Results are presented in Figure 3.e. One observes that the increase in humidity is accompanied by a linear augmentation of the radius of curvature from $R(S^X)$ = 4.9 ± 0.3 cm for RH = 9,5% to $R(S^X)$ = 5.8 ± 0.3 cm for RH = 86%. This suggests, according Stoney's theory, a residual stress modification into the silica matrix related to the chemical water adsorption. The variation of the curvature also indicates that the stress evolution is fully reversible upon adsorption/desorption cycles. The evolution of the thickness induced by water adsorption and desorption of the xerogel silica film deposited on silicon were assessed by *in situ* environmental ellipsometry and are reported in Figure 3.c. Here again, a reversible linear film transversal swelling is observed during the cycle of controlled humidity starting from 2% to 98%, without the presence of capillary contraction, confirming the lack of mesopores. The out-of-plane reversible strain of 1% highlights the release of a small part of the residual stress accumulated during the calcination step. The similar behavior of the curvature and of the film thickness with relative humidity evidences the isotropic relaxation of the residual stresses of the silica coating on polyimide substrate in humid atmosphere. Indeed, investigations on radius curvature provide direct information on longitudinal in-plane stress whereas environmental ellipsometry informs on stress perpendicular to the silica-substrate interface, the film being too thin for in plane variation due to mechanical consideration imposed by the rigid silicon substrate.[29]

Water seems thus to play the reversible role of stress sequestrating, as illustrated in Figure 5.a. By increasing the relative humidity, water molecules continuously adsorb onto the micropore surface until vapor saturation as depicted in Figure 5.a. Water molecule adsorption preferentially occurs on hydrophilic silanol-based groups, reducing thus the strength of the hydrogen bond between surface vicinal Si-OH groups. A part of the stress contained in silica matrix is thus transferred to water molecule through Si-OH/H$_2$O interaction leading to the decrease of the silica interfacial energy.

Because of the linear tendency, one may state that every adsorbed water molecule, interacting with the pore inner surface, participates to the interfacial energy transfer. It results in a progressive partial relaxation of the residual stress, as corroborated by the observed locale expansion of the microporous network.

At the bilayer strip scale (macroscopic), one can decompose the isotropic stress in three components: (i) the in-plane longitudinal component parallel to the interfaces, (ii) the out-of-plane component perpendicular to the interface, and (iii) the in-plane transversal component that is neglected due to the rectangular geometry of strip. The out-of-plane component is perpendicular to the silica-polyimide interface leading to the previously observed thickness swelling of the film. This strain phenomenon was first described by Bangham *et al* and Meehan on coal and is called Bangham's effect.[49,50] It describes the expansion of a materials induced by the adsorption of a wetting liquid due to the interfacial energy and the surface stress relaxations. In case of bilayer structure, the swelling of the xerogel silica film in adhesion with the polyimide substrate with low moisture sensitivity[51] creates a compression stress in opposition to the tensile residual stress, which decreases the bilayer curvature. The second in-plane component of the relaxation is in the same direction of the tensile residual stress. It probably takes directly part to the strip deformation, through a diminution of the silica film Young's effective moduli, perturbing the tensile force balance between the silica and the polyimide layers. The result is the reduction of the initial curvature of the strip. During desorption, the inversed process occurs, leading to the regeneration of the interfacial stresses, and thus to the initial curvature.

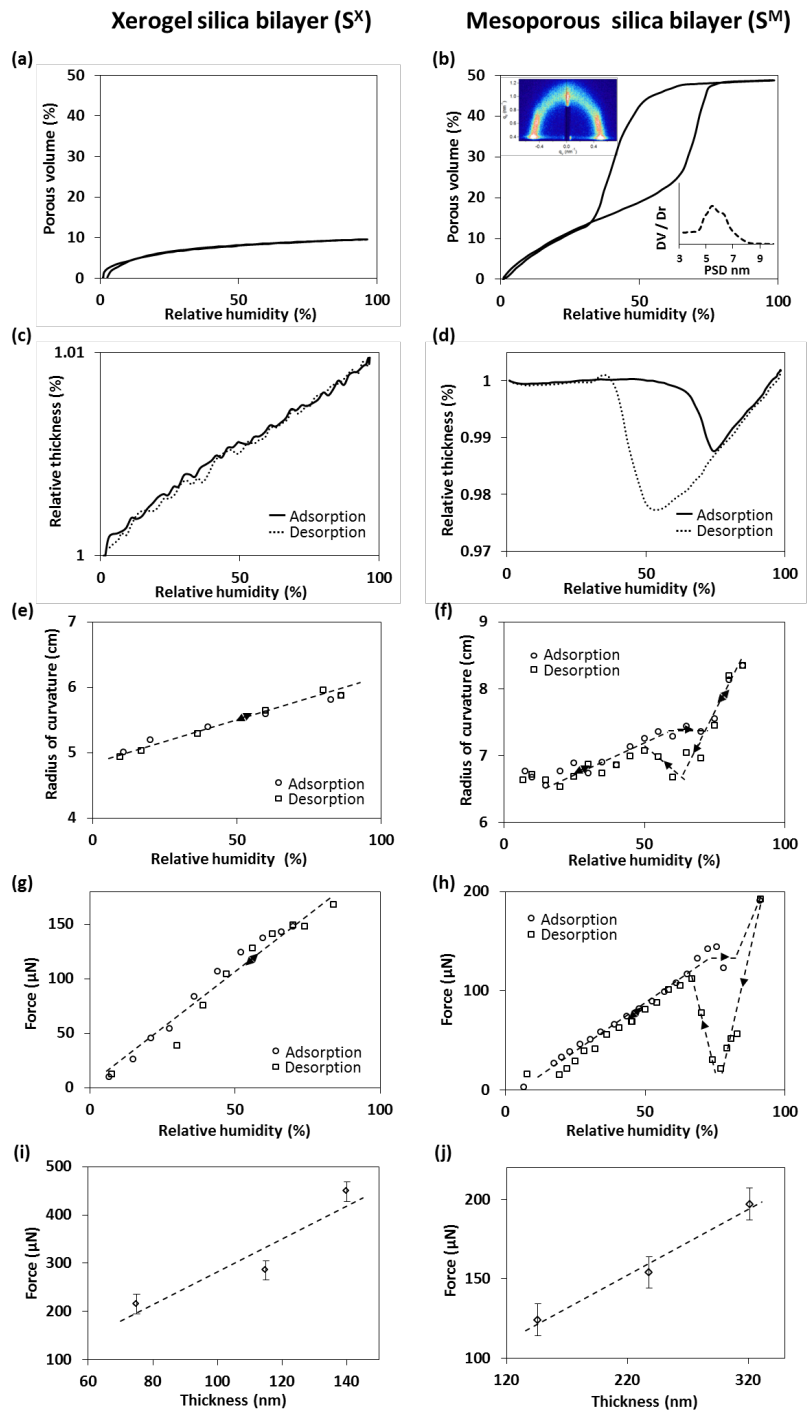

Figure 3. (a) and (b) Water adsorption-desorption isotherms of the microporous xerogel and the mesoporous F127-templated silica thin films on silicon substrates, with the corresponding pore size distribution, and the GISAXS pattern (inset) for the mesoporous thin film. (c) and (d) Relative thickness evolution depending on the relative humidity at room temperature for the microporous xerogel and the mesoporous F127-templated silica thin films on silicon substrates. (e) and (f) Radius of curvatures of the microporous xerogel ($S^X$) and the mesoporous ($S^M$) silica-polyimide bilayers strip depending on the relative humidity at room temperature. (g) and (h) Force generated by $S^X$ and $S^M$ actuators depending on the relative humidity at room temperature. (i) and (j) Values of force for microporous ($S^X$) and mesoporous ($S^M$) actuators versus the thickness of the silica layers at RH = 90%.

*Response of the F127-templated mesoporous thin film:*

The influence of water adsorption and desorption for the mesoporous silica responsive thin film and for the $S^M$ bilayer strips have been studied with the same approach than for the previous xerogel systems. Results are exposed in Figure 3.d and 3.f. The response of the mesoporous film is radically different from the xerogel one. A multistep increase of the radius of curvature is observed upon humidity raise with a first linear section from $R(S^M)$ = 6.5 ± 0.3 cm to $R(S^M)$ = 7.3 ± 0.3 cm between RH = 15% and RH = 55%. It is followed by a second section with stable values of $R(S^M)$ = 7.3 ± 0.3 cm between RH = 55% and RH = 72%. Finally the radius of curvature increases from $R(S^M)$ = 7.3 ± 0.3 cm to $R(S^M)$ = 8.3 ± 0.3 cm between RH = 72% and RH = 85%. The desorption is characterized by three steps. First, $R(S^M)$ changes from 8.3 ± 0.3 cm to $R(S^M)$ = 6.6 ± 0.3 cm between RH = 85% and RH = 62%. $R(S^M)$ increases back to 7.1 ± 0.3 cm down to RH = 50%. Between RH = 50% and RH = 15% $R(S^M)$ evolves from 7.1 ± 0.3 cm to $R(S^M)$ = 6.5 ± 0.3 cm. As for dense silica film, the mesoporous bilayer actuation is powered by the water affinity with silica. However, the whole behavior of the mesoporous film shows a hysteresis loop, in contrary to the xerogel film. Similarly, $S^M$ thickness evolution in Figure 3.d shows a hysteresis loop with two distinct capillary contractions due to the capillary adsorption and the capillary desorption of water into the pores, confirming the presence of mesopores. On the other hand, no significant thickness variations are observed at low pressures, suggesting that the Bangham's effect observed with the microporous thin film is hidden by the larger volume of mesopores, responsible for the dominant capillary contractions. It is likely that partial residual stress relaxation due to adsorption in the micropores, that are present into the inorganic network, takes place with a similar effect than the one observed for the microporous xerogel bilayer. It produces an in-plane relaxation, which induces the tensile stress decrease, resulting in the increase of the radius of curvature, as confirmed in Figure 3.f. But contrary to the xerogel films, no transversal swelling is observed, which is probably due to the presence of the mesopores that makes the network more flexible for isotropic relaxation (see Figure 5.b. step II). Indeed, the mesopore network is flexible enough to be able to accommodate the perpendicular relaxation due to the micropore "dilatation" induced by the adsorption, resulting in the "masking" of the transversal contribution of the micropore-induced-Bangham effect. Between 55% ≤ RH ≤ 70%, we observed the contraction of the mesoporous silica film on silicon substrate whereas the radius of curvature $R(S^M)$ keep constant for silica polyimide bilayer strips. This plateau of curvature seems to result from the cancelling of the stress relaxation by the capillary contraction of the film created by the capillary condensation. Over RH ≥ 70% both film thickness and radius of curvature significantly increase illustrating the recovering of surface stress relaxation of pores walls and pore curvatures. The higher slope of relaxation versus humidity compared to one observes for 7% ≤ RH ≤ 50% is probably due to the dissipation of stress accumulated during the capillary filling, illustrated in Figure 5.b. step III. During desorption, values of $R(S^M)$ do not take the same return path for the intermediate humidity (50% ≤ RH ≤ 70 %) due to the shift in vapor pressure at which desorption and adsorption take place (see figure 5.b step IV). The local minimal value of $R(S^M)$ = 6.6 ± 0.3 cm is associated to a breathing movement of the actuator, which comes from two combined phenomena, inducing both an abrupt contraction of the silica layer: (i) the intense stress generated by the isotropic capillary contraction and (ii) the residual stress recovery induced by the water desorption. Then for RH < 50% the bilayer strip recovers slowly its initial stress through the step by step water molecule desorption.

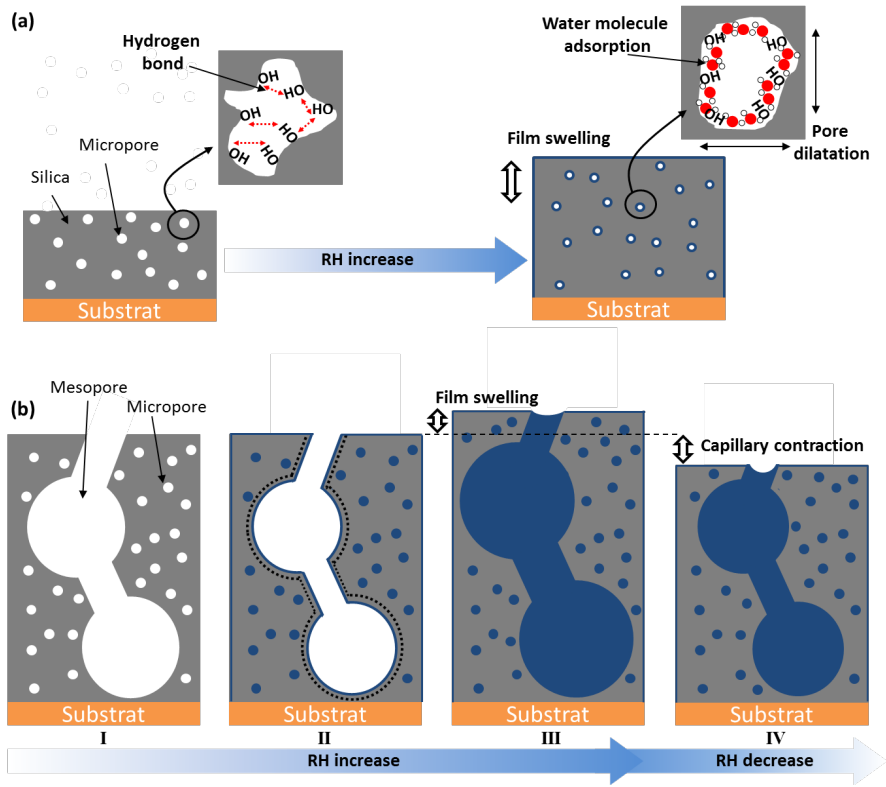

Figure 5. Scheme of the dynamic water adsorbed actuation mechanisms occurring in (a) microporous and (b) mesoporous responsive silica thin film for $S^X$ and $S^M$ actuators.

According to Stoney's equation, in-plane stress modulation of 15 % and 19 % were obtained for $S^X$ and $S^M$ actuators, respectively, when passing from RH = 10% to RH = 85%, highlighting a partial stress relaxation. It is clear from the latter results that RH could be used to actuate systems in combination with nanoporous inorganic thin films prepared by sol-gel. Increasing the surface area to volume ratio of silica films is probably the easiest way to enhance the mechanical movement of the silica-polyimide actuators by developing the surface of contact between water and silica, favoring the surface energy transfer from silica to adsorbat. One may also conclude that the adsorption of water at the surface and in the pores modifies the interfacial energy of the material and that the bond created at the interface may also relatively influence the strength of the bonds between atoms that are located underneath the interface inside the network.

*Evaluation and tuning of actuation force and speed*
In this part we evaluate the performance of the hybrid silica-polyimide actuators presented above. The effects of the pore size and the thickness of the responsive silica coatings on the force generated by the bilayer actuators have been deduced from *in situ* measurements of the force applied on a microforce sensing probe induced by the variation of the relative humidity in a sealed chamber (see Figure 6), and are reported in Figures 3.g and 3.h. As expected, the force globally builds up with the increase of humidity, similar to the variation of radius of curvature presented in Figure 3.e and 3.f. This shows that energy release at the water adsorption and at the capillary condensation can be converted in a useable, reversible, anisotropic force through a simple bilayer strip actuator.

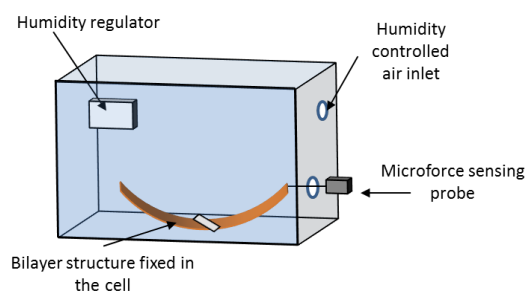

Figure 6. Representation of the set up used for the dynamic *in situ* investigation of the forces generated by the $S^X$ and $S^M$ bilayer actuators induced by the variation of the relative humidity.

Forces up to 200 µN could be generated with the actual device composition and dimension. As for the radius of curvature, the xerogel gives a fully reversible linear tendency, while the mesoporous has a more complex tendency with a linear tendency at low pressures accompanied by a hysteresis loop at high pressure that one can logically attribute to the capillary condensation. The shift of the hysteresis minimal values for the radius of curvature (Figure 3.f) compared to the measured force (Figure 3.h), does not have an evident rational explanation. It could however be attributed to a variation of temperature between the atmospheres of saturated vapor generation and of the measurement chamber. These very different force profiles indicate that this actuation can be used either for a progressive or a sudden conversion of energy in forces, depending on the pore size. The force can also be adjusted through the tuning of the porous film thickness, along which the stresses (tensile or compressive) and relaxation develop. The forces developed for a relative humidity of 90 % for various thicknesses are reported in Figure 3.i and 3.j for both types of coating combined with a 50 µm thick polyimide. As expected the developed forces increases quasi-linearly with the thickness of the sensitive coating, reaching 450 ± 20 µN and 200 ± 10 µN for a 140 ± 5 nm thick xerogel coating and for a 320 ± 5 nm thick mesoporous coating, respectively. First the linear augmentation suggests that the residual intrinsic stresses are proportionally distributed all along the silica thickness. It is worth noticing that values measured for the mesoporous actuators are significantly weaker than for the xerogel ones, despite the lower thickness and porosity of the latter. In other words, greater forces would be created with increasing specific surface area and thickness of the layer. Concerning the role of the polyimide support, force values of 910 ± 40 µN and 450 ± 20 µN were recorded at RH = 90% for bilayers strip composed a 140 nm thick xerogel layer deposited on a 75 µm thick and a 50 µm thick polyimide substrate, respectively. A thicker polyimide layer allowed developing a higher force through water adsorption due to a higher rigidity and also to its contribution on the stocked residual stress in the coating after thermal treatment (see Figure 2). In counterpart lower movement of amplitude was obtained for the thicker polyimide, confirming the classical compromise between the generated force and the amplitude of the movement

In addition a complete characterization of the actuator requires the study of actuation speed. A fast actuation speed is crucial to consider the integration of the actuator in a device. In table 1 are reported the speed of the deflection of actuators, in radius of curvature, depending on the relative humidity for xerogel ($S^X$) and mesoporous ($S^M$) silica films obtained from Figure 3.e and 3.f. Xerogel silica actuator, as seen previously, show a reversible linear increase of the radius of curvature with a constant speed of 1 %.%$^{-1}$ on all range of humidity. A similar magnitude of speed (i.e. 0.6 %.%$^{-1}$) for $S^M$ is observed at low pressures before capillary condensation. After capillary condensation a reversible linear increase of the radius of curvature is observed with a much higher speed value of 4 %.%$^{-1}$. The sudden increase of the humidity response of mesoporous actuators is similar to the voltage threshold for electronic diodes. Such mechanical actuators, showing an abrupt response with humidity threshold that is tunable through the size of pores, could be exploited as mechanical circuit breaker or switch.

Table 1. Relative speed of the radius of curvature variation versus the relative humidity depending on the range of relative humidity for S$^X$ and S$^M$ actuators.

|  | S$^X$ | S$^M$ |
|---|---|---|
| Speed 1 (%.%$^{-1}$) | 1 (9,5% ≤ RH ≤ 86%) | 0.6 (7% ≤ RH ≤ 50%) |
| Speed 2 (%.%$^{-1}$) | - | 4 (70% ≤ RH ≤ 85%) |

We demonstrated that reversible stress relaxation and capillary contraction inducing the movement of actuators was triggered by water molecule adsorption on the silica surface (mesopores, micropores). A second investigation of the actuation speed has been performed by measuring in parallel the force (using the microforce probe) and the relative humidity (using a lab hygrometer) as a function of the time. The mechanical response was always faster to stabilize than the hygrometer one, suggesting that, even if limited by the diffusion of gaseous water for both silica coating and hygrometer sensitive surface, the actuator shows probably a quasi-instantaneous response.

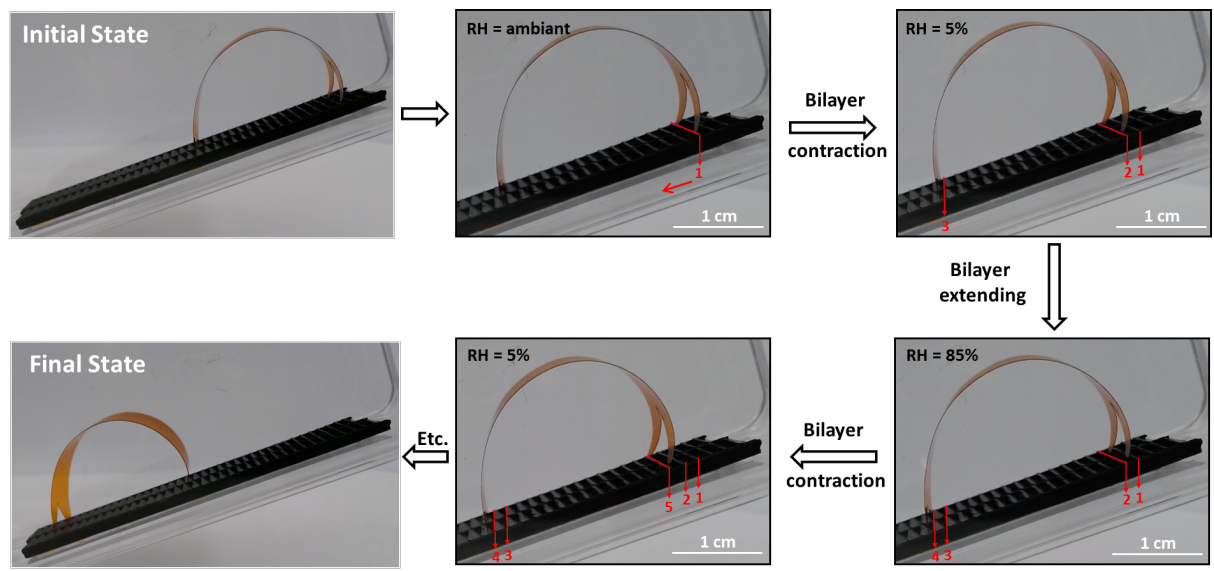

Figure 7. Illustration of the step-by-step moving of an bilayer actuator (50 µm thick polyimide layer combined to a 110 ± 5 nm xerogel silica film) placed on a tilted (≈15°) rack and pinion in humidity controlled chamber for humidity rate oscillating between 5% and 85%.

*The walking strip: an example of RH stimulated actuation.*
A bilayer actuator, optimized for large movement amplitude combined with high generated force of contraction and extension, has been obtained with a 110 ± 5 nm xerogel silica film deposited on a 50 µm thick polyimide layer. The curved actuator can stay standing-up on its 4 "legs" due to its relatively high rigidity but can perform a back and forth curvature variation under oscillating humidity. The actuator direction of displacement was directed by the use of a simple tilted rack with slightly inclined steps (see Figure 7). When a dry atmosphere is applied in the chamber, the actuator contracts inducing the jump of the upper legs from position 1 to position 2. Increasing the humidity induces the actuator extension and the jump of the lower legs from position 3 to position 4, the upper leg being blocked by the inclination. The repeated contractions and expansions of the actuator drive it down to the end of the rack (see movie available in supplementary information).

**Conclusions.**

It was shown that capillary condensation and/or water molecule adsorption in mesoporous and xerogel-based silica thin film could be converted in mechanical force using actuators with bilayer architecture to amplify the water induced nano-strain of a thin silica film backed on a thermos-resistant polyimide layer and provide at macroscopic scale useable mechanical work. We demonstrate that reversible surface energy reduction due to water adsorption, and Laplace forces induced during capillary condensation in nanopores, were responsible of the fast bilayer strip actuation. In view of the different responses obtained in this study, it is clear that one would be able to finely tune the mechanical response (speed, amplitude of movement, degree of force, force vs RH profile) by playing not only on the dimension of the pores, but on the porosity, on the morphology and orientation of the pores, on the interconnectivity between them, on the combination of various types of pores, without forgetting on the composition and the structure of the matrix itself. This fundamental work open the door to a new class of surface chemistry driven actuation systems implying non polymeric, but hard ceramic materials that could provide new solutions for energy harvesting devices or chemically driven muscle for robotics.

**Materials and Methods.**

Chemical reactants:
Absolute ethanol 100% (EtOH), TEOS (TetraEthylOrthoSilicate), hydrochloride acid 37%, F127 Pluronic ($EO_{106}PO_{70}EO_{106}$), were purchased from Aldrich. DuPont™ Kapton® HN sheets were purchased from Radiospares. All products were used as received.

Film processing:
Xerogel silica thin films, sample labeled $S^X$, were prepared from solutions composed of TEOS/HCl/$H_2O$/EtOH with respective molar ratio of 1:0.15:4.4:38. Mesoporous silica thin films with F127 (labeled $S^M$) template was prepared from solutions composed of TEOS/F127/HCl/$H_2O$/EtOH with respective molar ratios of 1:0.004:0.005:5:41.

TEOS was first dissolved in EtOH, HCl (2M) and $H_2O$ before potential addition of the template. Solutions were stirred at least for 24 h at room temperature before use. Films were prepared both on silicon wafer and 50 µm-thick Kapton substrate by dip-coating at room temperature and relative humidity of 10 ± 5% with constant withdrawal rates of 3, 7 or 12 mm.s$^{-1}$. After coating on Kapton, one face of the substrate was carefully cleaned with a tissue soaked with ethanol in order to remove the coating and keep a silica thin film on one side. Then a 60 mm long and 6 mm large strips were cut and immediately calcinated at 350°C for 10 min.

Characterization:
Spectroscopic ellipsometry and Environmental Ellipsometric Porosimetry (EEP) analyses were performed on a UV-visible variable angle spectroscopic ellipsometer (M2000 Woolam) equipped with a controlled atmosphere cell, in which relative vapor pressures of $H_2O$ was adjusted using mass flow controllers at room temperature (20 ± 2 °C).[31] Films thicknesses and refractive index were extracted from conventional Ψ and Δ dispersions using a Cauchy model (Wase 32 software).

For grazing incident small-angle X-ray scattering (GI-SAXS) a Rigaku S-max 3000 apparatus equipped with a microfocus source λ = 0.154 nm. The angle of incidence was 0.2°.

Curvature radius ($R$) were determined using ImageJ software from pictures of samples placed in a controlled atmosphere cell in which relative humidity was controlled by mass flow controllers

Microforce measurements were performed using a FemtoTools FT-S microforce sensing probe. Sample were fixed in humidity controlled chamber, the sensor was placed in contact with the bended end of the sample in perpendicular position. Compression force was measured between 50 nN and 1000 µN at 20Hz depending on the relative humidity.

## ASSOCIATED CONTENT

**Supporting Information**.
Video of the step-by-step moving of a bilayer actuator in controlled atmosphere.

## AUTHOR INFORMATION

**Corresponding Author**

* E-mail: david.grosso@upmc.fr.## ACKNOWLEDGMENT

M. Boudot *et al*. want to sincerely thanks Dr. Cédric Boissière and Dr. Marco Faustini for their fruitful discussion and help as well as M. Selmane for GI-SAXS experiments.**References.**